\documentstyle[epsfig,eqsecnum,aps]{revtex}  
\begin{document}
\preprint{submitted to PRA, Sept. 25, 2000}
\title{Quantum Magic Bullets via Entanglement}
\author{Seth Lloyd, Jeffrey H. Shapiro, and N. C. Wong}
\address{Research Laboratory of Electronics, 
Massachusetts Institute of Technology\\
Cambridge, Massachusetts 02139-4307}
\maketitle
\begin{abstract}
Two particles that are entangled with
respect to continuous variables such as position and
momentum exhibit a variety of nonclassical features. First,
measurement of one particle projects the other particle into
the state that is the complex conjugate of the state of the
first particle, i.e., measurement of one particle projects
the other particle into the time-reversed state. Second,
continuous-variable entanglement can be used to implement a
quantum ``magic bullet'': when one particle manages to pass
through a scattering potential, then no matter how low the
probability of this event, the second particle will also
pass through a related scattering potential with probability
one. This phenomenon is investigated in terms of the
original EPR state, and experimental realizations are
suggested in terms of entangled photon states.
\end{abstract}

\section{Introduction}
Entanglement is a peculiar quantum phenomenon in which two
quantum systems exhibit a greater degree of correlation than
is permitted classically. Entanglement has been shown to be
a highly useful effect for quantum computation and quantum
communications, see \cite{BennettShor} for a recent review. One of the
earliest and most striking pictures of entanglement was given by Einstein,
Podolsky, and Rosen (EPR) in their original paper on entangled states
\cite{EPR}. In contrast with the majority of subsequent work, the
original EPR paper concentrated on entanglement of
continuous variables, position and momentum. Although most
recent work on entanglement has focused on discrete systems
such as quantum bits or qubits, experimental demonstration
of continuous-variable teleportation via entanglement \cite{Furusawa},
theoretical constructions of analog quantum error-correcting codes
\cite{Braunstein1,Braunstein2,LloydSlotine}, and methods for
universal quantum computation over continuous variables
\cite{LloydBraunstein} suggest that the original EPR state is well-worth
revisiting.

This paper shows that entanglement of continuous variables
exhibits several significant properties in the context of
scattering theory. Consider two particles that are entangled
in terms of continuous variables such as position and
momentum.  Measurement of one particle can be shown to project
the other particle into the complex conjugate, or
time-reversed state. This fact has the following
consequence.  Consider a potential barrier put in the path of
the first particle that reflects some states and transmits
other states with probability one (such a barrier can be
thought of as a generalized filter). Then there is a related
potential barrier for the second particle---corresponding to
the time-reversed scattering matrix for the first potential---such that if
the first particle is transmitted through its barrier, the second particle
will be transmitted through its barrier with probability one. We call this
effect, a quantum ``magic bullet.'' This paper presents a theoretical
exposition of the magic bullet effect both in terms of the
original EPR state and in terms of entangled photons, and
proposes experimental realizations of quantum magic bullets.

\section{EPR-State Magic Bullets}
First, let us revisit the original Einstein-Podolsky-Rosen state.
Suppose that a particle with zero momentum decays into two
particles at time $t=0$. The particles have position eigenstates
$|x\rangle_j$, and momentum eigenstates
$|p\rangle_j = (1/\sqrt{2\pi\hbar})\int_{-\infty}^\infty
e^{ipx/\hbar}|x\rangle_j dx$, where $j=1,2$. (Like EPR, we
restrict our attention to a one-dimensional system: the
multi-dimensional generalization is straightforward.) Immediately after
the initial particle's decay, the joint state of the two new
particles is
\begin{equation}
|\psi\rangle_{\rm EPR} = \int_{-\infty}^\infty
|x\rangle_1|x\rangle_2\,dx = \int_{-\infty}^\infty
|p\rangle_1|-p\rangle_2\,dp.
\label{eq:EPR}
\end{equation}
These two particles are thus
perfectly correlated in position, and also perfectly
anticorrelated in momentum. It was this dual correlation
that Einstein, Podolsky and Rosen suggested was incompatible
with the classical notion of reality: the EPR state seems to
allow each particle to be in an eigenstate of two
noncommuting observables.

Irrespective of interpretations of quantum mechanics (which
are notoriously slippery), the EPR state exhibits the
following property. We expand $|\psi\rangle_{\rm EPR}$ in terms
of an arbitrary orthonormal basis $\{|\phi_z\rangle\}$, indexed by a
continuous parameter $z$, obtaining $|\phi_z\rangle_j =\int \phi_z(x)
|x\rangle_j dx$, where $\phi_z(x) = \langle x| \phi_z\rangle$.  Then,
recalling that $|x\rangle^*=|x\rangle$, we have
\begin{equation}
|\psi\rangle_{\rm EPR} =
\int |\phi_z\rangle_1|\phi_z\rangle^*_2\,dz.
\end{equation}
Suppose that a measurement of an operator with eigenstates
$|\phi_z\rangle_1 $ is made on the first particle. It is
immediately seen that if this measurement reveals the first
particle to be in the state $|\phi_z\rangle_1$, the second
particle is then in the complex conjugate state
$|\phi_z\rangle^*_2$.

Complex conjugation in quantum mechanics is equivalent to
time reversal. If a particle evolves according to a
Hamiltonian $\hat{H}$, so that an initial state $|\phi\rangle$ at $t=0$
evolves into the state $|\phi(t)\rangle=
e^{-i\hat{H}t/\hbar}|\phi\rangle$ for $t>0$, then taking the complex conjugate
gives the state $e^{i\hat{H}^*t/\hbar}|\phi\rangle^*$, viz., the complex
conjugate of the evolved state is the complex conjugate of
the initial state evolved backward in time ($t\rightarrow
-t$) according to the time-reversed Hamiltonian $\hat{H}^*$. If
$\hat{H}$ is real, i.e., time-reversal invariant (as is the case
for almost all fundamental Hamiltonians, with the notable
exception of the $K^0$ meson), then the energy
eigenfunctions of $\hat{H}$ can be taken to be real, because
$\hat{H}|E\rangle = E|E\rangle$ implies $\hat{H}|E\rangle^* =
E|E\rangle^*$. Suppose that both particles in the EPR state
are subjected to a time-reversal invariant Hamiltonian $\hat{H}$.
Over time $t$, the state $|\psi\rangle_{\rm EPR}$ evolves into
\begin{equation}
|\psi(t)\rangle_{\rm EPR} \equiv \int e^{-i\hat{H}t/\hbar} |\phi_z\rangle_1
e^{-i\hat{H}t/\hbar}|\phi_z\rangle^*_2\,dz = \int
|\phi_z(t)\rangle_1|\phi_z(-t)\rangle^*_2\,dz.
\end{equation}
The state of the first particle at
time $t$ is thus perfectly correlated with a state that is the complex
conjugate of that particle's state at time $-t$: the second particle is
in the time-reversed state of the first.

The EPR state also exhibits interesting scattering
properties. Suppose that the first particle is subjected to
a time-reversal invariant Hamiltonian corresponding to a
unitary scattering matrix $\hat{S}(t)=e^{-i\hat{H}t/\hbar}$. Note that
$\hat{S}^*(t)=\hat{S}^\dagger(t)=\hat{S}^{-1}(t)$. After the particle is
scattered, let us make a projective measurement on it, corresponding to the
Hermitian operator $\hat{O}=\sum_o o \hat{P}_o$, where $o$ is real and
$\hat{P}_o=
\hat{P}^2_o$ is the projection operator on the eigenspace
corresponding to $o$. Now suppose that the second particle is
subjected to the inverse scattering operation $\hat{S}^*(t)$, and that the
conjugate measurement $\hat{O}^* =\sum_o o\hat{P}_{o}^*$ is made on this 
particle. Then the results
of the two measurements will be perfectly correlated: a
result of $o$ for the first particle will be accompanied by
a result $o$ for the second particle. That is, if the first
particle passes through a given scattering potential, the
second particle passes through the time-reversed potential
with probability one, regardless of how unlikely it was for the first
particle to have breached its barrier.  We call this effect a
quantum magic bullet.

The magic bullet effect can be implemented in other ways as
well. For example, suppose that it is possible to conjugate
the phase of the second particle (for example, by sending it
off a phase-conjugate mirror) at time $t$. The phase
conjugation effectively performs a spin echo on the
particle, resulting in the state $\int
|\phi_z(t)\rangle_1|\phi_z(-t)\rangle_2\,dz$. The second
particle now performs the same dynamics as the first
particle, but with a time lag of $2t$. In particular, if the
first particle manages to pass through a generalized filter,
then the second particle will pass through the same filter
with probability one $2t$ seconds later.

Quantum magic bullets can have many manifestations. We now
turn to examples of quantum magic bullets that can
be constructed using nonlinear optics.

\section{Field-Quadrature Magic Bullets}
Parametric interactions in $\chi^{(2)}$ crystals have proved to be rich
sources of quantum light-beam phenomena, see
\cite{ShapiroSun,ShapiroWong} for a unified treatment of a wide
variety of such effects, including quadrature-noise squeezing, nonclassical
twin-beam production, nonclassical fourth-order interference, and
polarization-entangled photon-pair production.  All these phenomena
originate from the same fundamental physics:  in a $\chi^{(2)}$ material
pumped by a strong beam at frequency $\omega_P$ and wave vector ${\bf
k}_P$, a single pump photon is converted into a pair of photons---one signal
($S$) and one idler ($I$)---subject to the energy- and
momentum-conservation conditions, i.e., $\omega_S + \omega_I = \omega_P$ and
${\bf k}_S + {\bf k}_I = {\bf k}_P$, respectively.

We shall use the
continuous-wave, type-II phase matched, doubly-resonant optical parametric
amplifier (OPA)---with vacuum-state signal and idler inputs---as the basis
for all of the optical magic-bullet realizations to be discussed.  This OPA
arrangement, shown schematically in Fig.~1, produces signal and idler
outputs with  orthogonal polarizations, well-defined spatial modes, and
fluorescence bandwidths in the MHz to GHz range.  As in \cite{ShapiroWong},
we shall assume that the signal and idler linewidths are identical, that
there are no losses in the OPA cavity, and that there is no depletion of nor
excess noise on the pump beam.  The positive-frequency, photon-units field
operators for the excited output polarizations of the signal and idler,
$\hat{E}_{S}(t)$ and $\hat{E}_{I}(t)$, are then conveniently
expressed in terms of their respective fluorescence center frequencies,
$\omega_{S}$ and $\omega_{I}$, and their complex envelopes via,
$\hat{E}_j(t) = \hat{A}_j(t)e^{-i\omega_jt}$, for $j = S, I$.
The full, multi-mode, joint state of these output signal and idler
fields is known to be a stationary, entangled, Gaussian pure
state that is completely characterized by the following normally-ordered
(fluorescence) and phase-sensitive spectra \cite{ShapiroWong}:
\begin{eqnarray}
S^{(n)}(\omega) &\equiv& \int_{-\infty}^{\infty}
\langle \hat{A}_{S}^\dagger(t+\tau)\hat{A}_{S}(t)\rangle
e^{-i\omega\tau}\,d\tau   \nonumber\\
&=& \int_{-\infty}^{\infty}
\langle \hat{A}_{I}^\dagger(t+\tau)\hat{A}_{I}(t)\rangle
e^{-i\omega\tau}\,d\tau =
\left|\frac{\displaystyle 2G}
{\displaystyle 1 - G^2 -(\omega/\Gamma)^2 - 2i\omega/\Gamma}\right|^2,
\label{eq:normspect}\\[.12in]
S^{(p)}(\omega) &\equiv& \int_{-\infty}^{\infty}
\langle \hat{A}_{S}(t+\tau)\hat{A}_{I}(t)\rangle
e^{-i\omega\tau}\,d\tau =
\frac{\displaystyle 2G[1 + G^2 +(\omega/\Gamma)^2]}
{\displaystyle |1 - G^2 -(\omega/\Gamma)^2 - 2i\omega/\Gamma|^2}.
\label{eq:phasespect}
\end{eqnarray}
Here, $G^2$ is the OPA pump power, normalized to the threshold power for
oscillation, and $\Gamma$ is the cavity-loss rate.

To establish an analogy between the multi-mode signal and idler fields and
the two-particle EPR state, let us consider an arbitrary pair of entangled
single-frequency modes, namely, the signal beam at frequency $\omega_S +
\Delta\omega$ and the idler beam at frequency $\omega_I - \Delta\omega$.
The joint signal$\times$idler state for these two modes has the
number-ket representation,
\begin{equation}
|\psi\rangle_{SI} = \sum_{n=0}^{\infty}
\sqrt{\frac{\displaystyle \bar{N}^{n}}{\displaystyle (\bar{N}+1)^{n+1}}}\,
|n\rangle_{S}|n\rangle_{I},
\label{eq:BoseEinstein}
\end{equation}
where $\bar{N} = S^{(n)}(\Delta\omega)$ is the average number of photons
per mode.  Individually, each
mode (signal and idler) is in a chaotic (Bose-Einstein) state, but their
photon numbers are perfectly correlated.  We shall return to this photon-pair
property in the sections to follow.  Our present course is to connect this
signal$\times$idler state to the EPR state.  For that purpose we need the
field-quadrature representation for $|\psi\rangle_{SI}$.

The real and imaginary parts of a photon annihilation operator, $\hat{a}$,
i.e, its quadrature components $\hat{a}_1 \equiv \mbox{Re}(\hat{a})$ and
$\hat{a}_2 \equiv \mbox{Im}(\hat{a})$, behave like normalized
versions of position and momentum.  In particular, the $\hat{a}_1$
eigenkets, $|\alpha_1\rangle$, are related to the $\hat{a}_2$ eigenkets,
$|\alpha_2\rangle$, by Fourier transformation, $|\alpha_2\rangle =
(1/\sqrt{\pi})
\int_{-\infty}^{\infty}e^{2i\alpha_2\alpha_1}|\alpha_1\rangle\,d\alpha_1$.
The joint signal$\times$idler state given in Eq.~\ref{eq:BoseEinstein} takes
the following form, when written in the field-quadrature representation
generated by the eigenkets of $\hat{a}_{S_{1}}$ and $\hat{a}_{I_{1}}$,
\begin{equation}
|\psi\rangle_{SI} = \int_{-\infty}^{\infty}\int_{-\infty}^{\infty}
\psi(\alpha_{S_{1}},\alpha_{I_{1}})|\alpha_{S_{1}}\rangle_{S}
|\alpha_{I_{1}}\rangle_{I}\,d\alpha_{S_{1}}\alpha_{I_{1}},
\label{eq:quadentangle1}
\end{equation}
with
\begin{equation}
\psi(\alpha_{S_{1}},\alpha_{I_{1}}) \equiv
\exp\!\left[-(1+2\bar{N})\alpha_{S_{1}}^{2} +
4\sqrt{\bar{N}(\bar{N}+1)}\,\alpha_{S_{1}}\alpha_{I_{1}} -
(1+2\bar{N})\alpha_{I_{1}}^{2}\right]/\sqrt{\pi/2}.
\label{eq:quadentangle2}
\end{equation}

Equations~\ref{eq:quadentangle1} and \ref{eq:quadentangle2} are not
identical to the EPR state, Eq.~\ref{eq:EPR}, but they do embody a
nonclassical continuous-variable correlation.  Optical homodyne detection
\cite{YuenShapiro} can be used to perform the $\hat{a}_{S_{1}}$ and
$\hat{a}_{I_{1}}$ measurements.  When such measurements are made on
the state $|\psi\rangle_{SI}$, the unconditional (marginal) statistics
for the $\hat{a}_{S_{1}}$ and $\hat{a}_{I_{1}}$ observations are
identical:  their individual outcomes are Gaussian random variables, each
with mean zero and variance $(1+2\bar{N})/4$.  It is the joint
statistics of these two measurements that reveals pure quantum
behavior.  In particular, if we are given that the outcome of the
$\hat{a}_{S_{1}}$ measurement is $\alpha_{S_{1}}$, then the
conditional statistics of the $\hat{a}_{I_{1}}$ measurement remain
Gaussian, but with conditional mean
$[\sqrt{4\bar{N}(\bar{N}+1)}/(1+2\bar{N})]\alpha_{S_{1}}$ and
conditional variance
$1/4(1+2\bar{N})$.  Thus, when $\bar{N}\ge 1$ there is a strong
sub-shot-noise correlation between these signal-beam and idler-beam homodyne
measurements:  the conditional variance is substantially below that
coherent-state (shot-noise) level of 1/4.  Moreover, as
$\bar{N}\longrightarrow\infty$, the state $|\psi\rangle_{SI}$ approaches
a normalized version of the EPR state, as can be seen from rewriting
$\psi(\alpha_{S_{1}},\alpha_{I_{1}})$ as follows:
\begin{eqnarray}
\psi(\alpha_{S_{1}},\alpha_{I_{1}}) &=&
\frac{\displaystyle \exp\!\left(-\frac{\displaystyle \alpha_{{S}_1}^2}
{\displaystyle 1+2\bar{N}}\right)}{\displaystyle [\pi(1+2\bar{N})/2]^{1/4}}
\frac{\displaystyle \exp\!\left[-(1+2\bar{N})
\left(\alpha_{{I}_1} - \frac{\displaystyle \sqrt{4\bar{N}(\bar{N}+1)}}
{\displaystyle 1+2\bar{N}}\alpha_{{S}_1}\right)^2
\right]}{\displaystyle [\pi/2(1+2\bar{N})]^{1/4}} \\[.12in]
&\longrightarrow& \frac{\displaystyle 1}{\displaystyle (\pi\bar{N})^{1/4}}
\delta(\alpha_{I_{1}}-\alpha_{S_{1}}),\quad
\mbox{as $\bar{N}\longrightarrow\infty$.}
\end{eqnarray}

Experimental demonstrations of this signal/idler homodyne correlation have
already been reported \cite{Reynaud,Leong}, although for optical
parametric oscillators (OPOs)---in which the strong mean fields arising from
above-threshold OPA operation act as homodyne-detection local
oscillators---rather than for OPAs.  An example of such data, obtained from
a triply-resonant optical parametric oscillator \cite{Teja}, is shown in
Fig.~2.  The upper trace shows the shot-noise level and the lower trace
shows the signal-minus-idler intensity difference for frequency detunings,
$\Delta\omega/2\pi$, ranging from 2 to 6\,MHz.  The $>5$\,dB noise reduction
in the signal-minus-idler intensity difference is a manifestation of the
nonclassical correlation cited above for the OPA field quadratures.

\section{Photon-Pair Magic Bullets}
Most entanglement experiments that rely on parametric optical interactions
draw upon the photon-pair property exhibited in Eq.~\ref{eq:BoseEinstein}.
Moreover, these experiments---which employ non-resonant, parametric
downconverters rather than doubly-resonant optical parametric
amplifiers---are carried out at extremely low photon fluxes.  In this
regime, a $T$-sec-long photon-counting measurement (on either the signal or
idler beam) will yield zero counts with near-unity probability $1-p$, and one
count with probability $p$; the probability of multiple counts at such low
fluxes is negligible.  Photon-pair creation within the $\chi^{(2)}$ medium
is nearly instantaneous, but, for our doubly-resonant OPA, the time
correlation between signal and idler photons in the output beams is smeared
out to several cavity lifetimes.  Thus, to represent the OPA version of
low-flux photon-pair generation,  we decompose
$\hat{E}_{S}(t)$ and
$\hat{E}_{I}(t)$---on a photon-counting interval $[0, T]$, where $\Gamma
T\gg 1$---into operator-valued Fourier series whose coefficients are the
photon annihilation operators,
\begin{eqnarray}
\hat{a}_{{S}_n} &\equiv&
\int_{0}^{T}\hat{E}_{S}(t)
\frac{\displaystyle \exp[i(\omega_{S} + 2\pi n/T)t]}{\displaystyle
\sqrt{T}}\,dt,\\[.12in]
\hat{a}_{{I}_n} &\equiv&
\int_{0}^{T}\hat{E}_{I}(t)
\frac{\displaystyle \exp[i(\omega_{I} - 2\pi n/T)t]}{\displaystyle
\sqrt{T}}\,dt.
\end{eqnarray}
When there is one photon
pair present in $[0,T]$, its joint state then has the entangled
multi-mode number-ket expansion,
\begin{equation}
|\psi\rangle_{SI} =
\sum_{n}\psi_n|1\rangle_{{S}_n}|1\rangle_{{I}_n},
\end{equation}
in this representation,
where $|\psi_n|^2 \propto S^{(n)}(2\pi n/T)$, and our Fourier-decomposition
sign convention has forced there to be an $\hat{a}_{I_{n}}$ photon
present whenever an
$\hat{a}_{S_{n}}$ photon occurs.  From this photon-pair state we can
exhibit the conjugate-state projection property of the magic bullet effect.

Suppose that we make a measurement that projects the signal photon onto the
state
$|\phi\rangle_{S} = \sum_{n}\phi_n|1\rangle_{S_{n}}$.  When that
measurement yields a non-zero result, it is easy to see that the idler
photon is left in the state,
\begin{equation}
|\psi\rangle_{I} =
\frac{\displaystyle \sum_{n}\psi_n\phi_n^*|1\rangle_{I_{n}}}
{\displaystyle \sqrt{\sum_{n}|\psi_n|^2|\phi_n|^2}}.
\end{equation}
If the $\psi_n$ are approximately constant over the $n$ values for which
$|\phi_n|$ differs significantly from zero, then (except for a physically
insignificant absolute phase) we get
\begin{equation}
|\psi\rangle_{I} \approx \sum_n \phi_n^*|1\rangle_{I_{n}},
\end{equation}
i.e., the projective measurement on the signal photon has placed the idler
photon in the conjugate state.

To our knowledge, the preceding conjugate-state projection property of
entangled photon pairs has not been observed experimentally.  It does,
however, have an important application in quantum communications.  The
recently proposed system for long-distance entanglement transmission
(through standard telecommunication fiber) and long-duration optical storage
(in trapped-atom quantum memories) \cite{Shapiro} implicitly uses this
effect.  Let us make that cavity effect explicit.

Consider two high-$Q$,
initially unexcited, single-ended optical cavities---resonant at frequencies
$\omega_{S} +
\Delta\omega$ and $\omega_{I}  - \Delta\omega$, respectively---that have
no excess losses.  Suppose that these cavities are illuminated by the signal
and idler fields $\hat{E}_{S}(t)$ and $\hat{E}_{I}(t)$ for a
$T_c$-sec-long time interval, after which photon-counting measurements
$\hat{n}_{S} \equiv
\hat{a}_{S}^\dagger(T_c)\hat{a}_{S}(T_c)$ and $\hat{n}_{I}
\equiv
\hat{a}_{I}^\dagger(T_c)\hat{a}_{I}(T_c)$ are performed on the
resulting cavity fields.  The intracavity photon annihilation operators at
time $T_c$, $\hat{a}_{S}(T_c)$ and
$\hat{a}_{I}(T_c)$, are related to the initial (vacuum-state) cavity
operators, $\hat{a}_{S}(0)$ and
$\hat{a}_{I}(0)$, and the input signal and idler fields via,
\begin{eqnarray}
\hat{a}_{S}(T_c) &=& \hat{a}_{S}(0)
e^{-(\Gamma_c + i\Delta\omega)T_c} +
\int_{0}^{T_c}\!dt\,\sqrt{2\Gamma_c}
e^{-(\Gamma_c+ i\Delta\omega)(T_c-t) + i\omega_{S}t}\hat{E}_{S}(t),\\[.12in]
\hat{a}_{I}(T_c) &=& \hat{a}_{I}(0)
e^{-(\Gamma_c - i\Delta\omega)T_c} +
\int_{0}^{T_c}\!dt\,\sqrt{2\Gamma_c}e^{-(\Gamma_c - i\Delta\omega)(T_c-t) +
i\omega_{I}t}
\hat{E}_{I}(t),
\end{eqnarray}
where $\Gamma_c$ is the measurement-cavity linewidth.
The OPA statistics from
Eqs.~\ref{eq:normspect} and \ref{eq:phasespect} can now be used to evaluate
the normalized photocount-difference variance,
$\sigma^{2}_{n} \equiv \langle(\hat{n}_{S} -
\hat{n}_{I})^2\rangle/(\langle\hat{n}_{S}\rangle +
\langle\hat{n}_{I}\rangle)$, from which the presence of the
magic-bullet effect can be deduced.

The magic-bullet effect occurs in the narrowband measurement
regime ($\Gamma_c \ll \Gamma$), when the cavity-loading time
$T_c$ is long enough for statistical steady-state to be reached
($\Gamma_c T_c \gg 1$). In Fig.~3 we have plotted $\sigma^2_{n}$ vs.\
$\Gamma_c/\Gamma$, for several values of the normalized detuning,
$\Delta\omega/\Gamma$, with $G^2 = 0.01$ and  $\Gamma_cT_c\gg
1$.  Figure~3 clearly shows the magic bullet effect as $\Gamma_c/\Gamma
\longrightarrow 0$.  The low photon-fluxes of the signal and idler beams
imply that the $\hat{n}_{S}$ and $\hat{n}_{I}$ measurements each
yield outcomes that are either zero or one.  For there to be a strongly
sub-shot-noise value of the normalized photocount-difference variance
($\sigma^2_{SI}\ll 1$), it must be that every signal-cavity count is
accompanied by an idler-cavity count, even though the probability that a
signal photon will make it into its cavity becomes very low as
$\Gamma_c/\Gamma$ decreases.  Note that
$\Gamma_c/\Gamma\ll 1$ makes the signal and idler fluorescence spectra
approximately constant over their respective measurement-cavity linewidths,
in keeping with the general conjugate-state projection requirement that the
$\psi_n$ be approximately constant over the $n$ values for which $|\phi_n|$
differs significantly from zero.

\section{Magic-Bullet Penetration of Optical Filters}
The cavity-loading conjugate-state projection example that we have just seen
extracts single-mode measurements---the photon number in each cavity---from
the multi-mode illumination fields, $\hat{E}_{S}(t)$ and $\hat{E}_{I}(t)$.
Our final example of the quantum magic-bullet effect will examine
multi-mode measurements of these multi-mode fields.

Suppose that the signal and idler beams from the OPA illuminate a pair of
narrowband optical-transmission filters, and that the outputs from these
filters, in turn, illuminate a pair of unity quantum efficiency
photodetectors.  As shown in Fig.~4, we will assume that these filters are
symmetrically displaced from the center frequencies of the signal and idler
fluorescence spectra, so as to select frequency pairs that are entangled.  A
magic-bullet effect, if it exists in this framework, would involve photon
counting over $T$-sec-long time intervals satisfying $\omega_c T\gg 1$,
where $\omega_c$ is the filter bandwidth.  In particular, using
$\hat{N}_{S}$ and $\hat{N}_{I}$ to denote the signal and idler count
measurements over these intervals, low-flux OPA operation implies that these
measurements each yield either zero counts (a high-probability event) or one
count (a low probability event).  Moreover, the probability that any
particular signal photon will successfully pass through the narrowband
signal-beam filter will be very low when $\omega_c\ll \Gamma$.  Thus, if
there is a strongly
sub-shot-noise signal-minus-idler normalized photocount-difference
variance, $\sigma^{2}_{N} \equiv \langle(\hat{N}_{S} -
\hat{N}_{I})^2\rangle/(\langle\hat{N}_{S}\rangle +
\langle\hat{N}_{I}\rangle) \ll 1$, it must be that whenever a signal
photon is transmitted by the signal-beam filter, there is an accompanying
idler photon that is transmitted by its filter.  This
signal-transmission/idler-transmission pairing is the magic-bullet effect:
it occurs whenever $\sigma^{2}_{N}\ll 1$, regardless of how unlikely it is
for a signal photon to pass through the signal-beam filter.

Magic-bullet filter penetration is intrinsically a multi-mode effect,
because of the large time-bandwidth product, $\omega_c T \gg 1$, that is
involved.  To analyze this situation, let us assume that the signal and idler
filters have no excess losses, and that their intensity
transmissions have
$K$th-order Butterworth shapes given by,
\begin{eqnarray}
|H_{S}(\omega)|^{2} &=&
\frac{\displaystyle 1}
{\displaystyle 1 + [(\omega-\omega_{S} - \Delta\omega)/\omega_c]^{2K}},
\\[.12in]
|H_{I}(\omega)|^{2} &=&
\frac{\displaystyle 1}
{\displaystyle 1 + [(\omega-\omega_{I} + \Delta\omega)/\omega_c]^{2K}}.
\end{eqnarray}
The OPA statistics from \cite{ShapiroWong} can be combined with the analysis
techniques from \cite{ShapiroSun} to show that
$\sigma^2_N \approx 1/2K$,
for $\omega_c/\Gamma\ll 1$ when $\omega_c T\gg 1$. Evidently, there is a
magic bullet effect here, but it requires the use of steep-skirted optical
filters, i.e., $K\gg 1$.
\section{Discussion}
In this paper we have laid out the basic properties of quantum magic
bullets.  Starting from the continuous-variable entanglement considered by
Einstein, Podolsky, and Rosen, we have shown that a projective measurement
on one particle of an entangled pair projects the other into the conjugate
(time-reversed) state.  Conjugate-state projection, in turn, permitted us to
show that when one particle successfully negotiates a scattering potential,
its entangled companion will pass through a related scattering potential with
probability one, no matter how unlikely the first event was.  This is
the quantum magic bullet.  In seeking optical realizations of magic bullets,
we first showed that the field-quadrature entanglement that exists between
appropriately-paired signal and idler frequencies from optical
parametric interactions approximates the EPR-state.  The sub-shot-noise
levels seen in OPA quadrature-squeezing experiments \cite{Wu} implicitly
confirm this behavior.  The photon-twins behavior seen in OPO
intensity-difference measurements
\cite{Reynaud,Leong,Teja}, provide a direct demonstration of
the nonclassical correlation between the signal and idler frequencies.  The
full, joint Bose-Einstein state has also been seen, in the output from a
parametric downconverter, via quantum-state tomography \cite{Kumar}.

The field-quadrature form of optical magic bullets is an asymptotic
effect that is strongly nonclassical only when the average photon number
per mode is high.  Furthermore, it requires the use of homodyne or
self-homodyne measurements.  Photon-pair counting measurements in the
low-flux operating regime provide a more attractive optical magic-bullet
scenario, although they do not represent the perfect analog of the EPR-state
position-momentum entanglement that the $\bar{N}\longrightarrow \infty$
field-quadrature state does.  The single-mode realization of the photon-pair
magic bullet, based on intracavity photon-counting, has yet to be
demonstrated experimentally.  Nevertheless, it is intrinsic to the coupling
of polarization-entangled photons from an OPA pair \cite{ShapiroWong} into a
trapped-atom quantum memory \cite{LloydShahriar} for the purpose of
long-distance transmission and long-duration storage of qubits
\cite{Shapiro}.  The multi-mode realization of the photon-pair magic bullet
requires that steep-skirted, low-excess-loss filters be used.  A possible
experimental realization might use a grating/lens/pinhole system in which
the frequency components of the signal and idler beams were first dispersed
in angle, then focused to spatially separate them on a detector plane, where
a pinhole would provide steep-skirted frequency selection prior to
photodetection.

\acknowledgments

This research was supported by the National Reconnaissance Office under contract 
NRO000-00-C-0032.
\newpage
\begin{figure}
\vspace*{0.5in}
\begin{center}
\epsfig{file=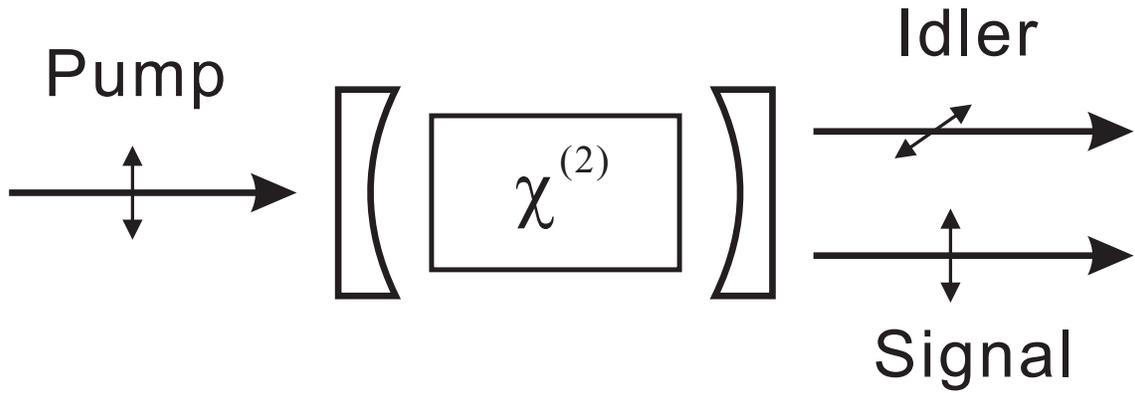,width=15cm}
\end{center}
\caption{Schematic of a doubly-resonant optical 
parametric amplifier.}
\end{figure}

\begin{figure}
\vspace*{0.5in}
\begin{center}
\epsfig{file=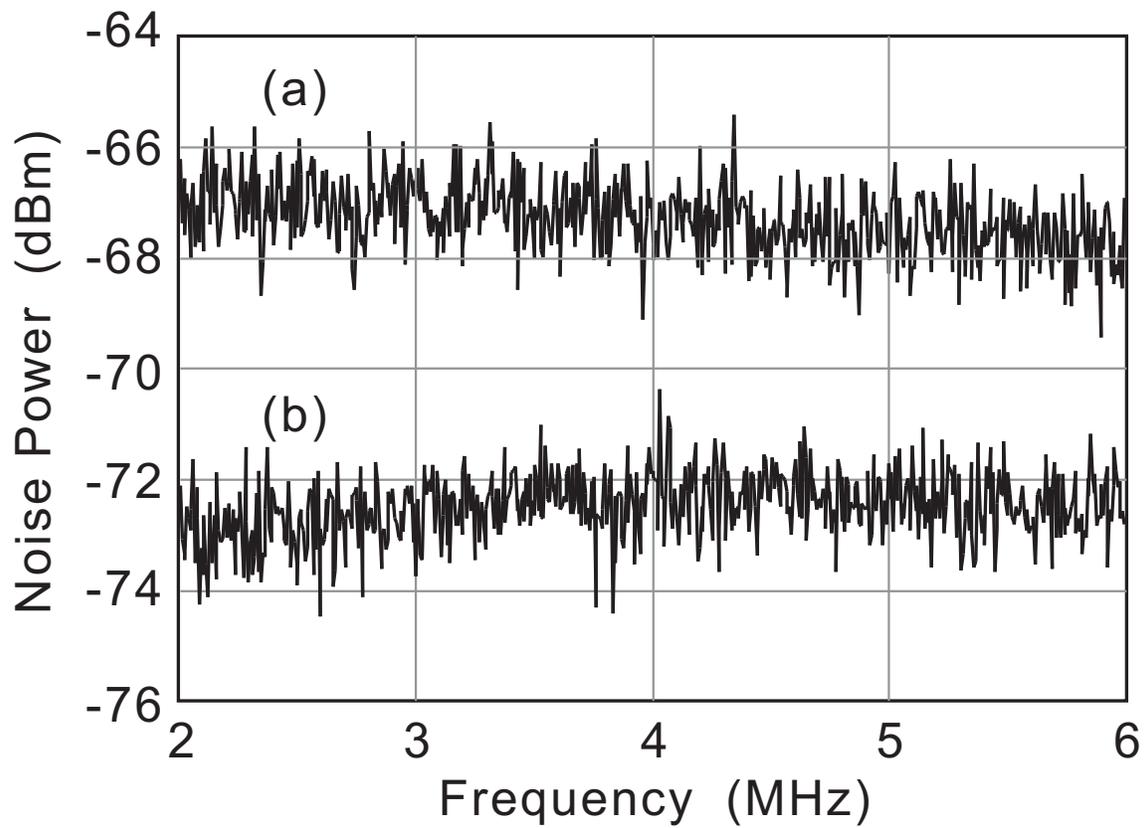,width=15cm}
\end{center}
\caption{Shot-noise level (a), and 
signal-minus-idler intensity difference (b), from a KTP optical parametric 
oscillator.}
\end{figure}

\begin{figure}
\vspace*{0.5in}
\begin{center}
\epsfig{file=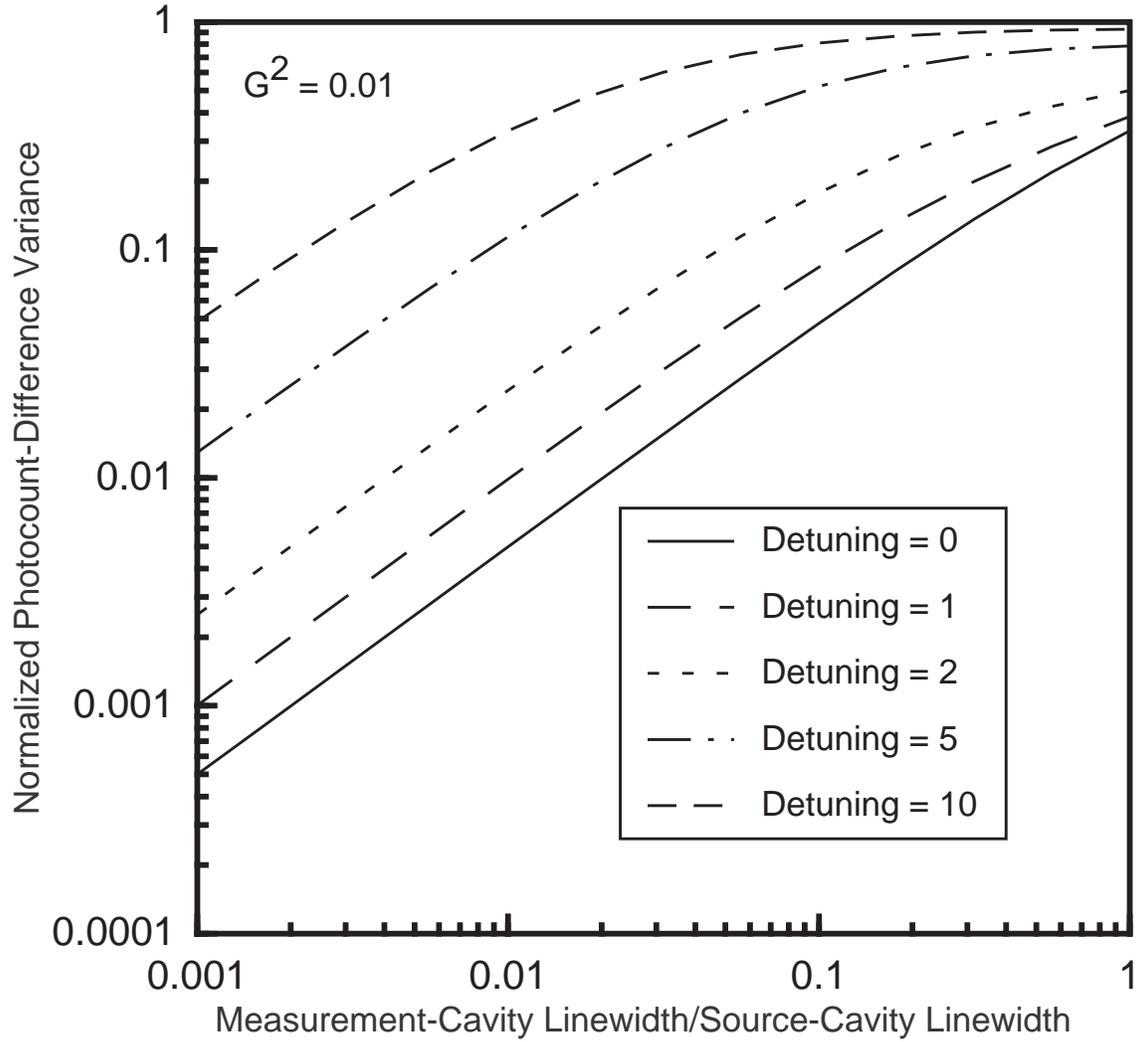,width=15cm}
\end{center}
\caption{Normalized photocount-difference 
variance, $\sigma^2_n$, vs.\
the ratio of measurement-cavity linewidth to source-cavity linewidth,
$\Gamma_c/\Gamma$, for 1\% OPA pumping ($G^2 = 0.01$) and various
values of the normalized detuning, $\Delta\omega/\Gamma$.}
\end{figure}

\newpage
\begin{figure}
\vspace*{0.5in}
\begin{center}
\epsfig{file=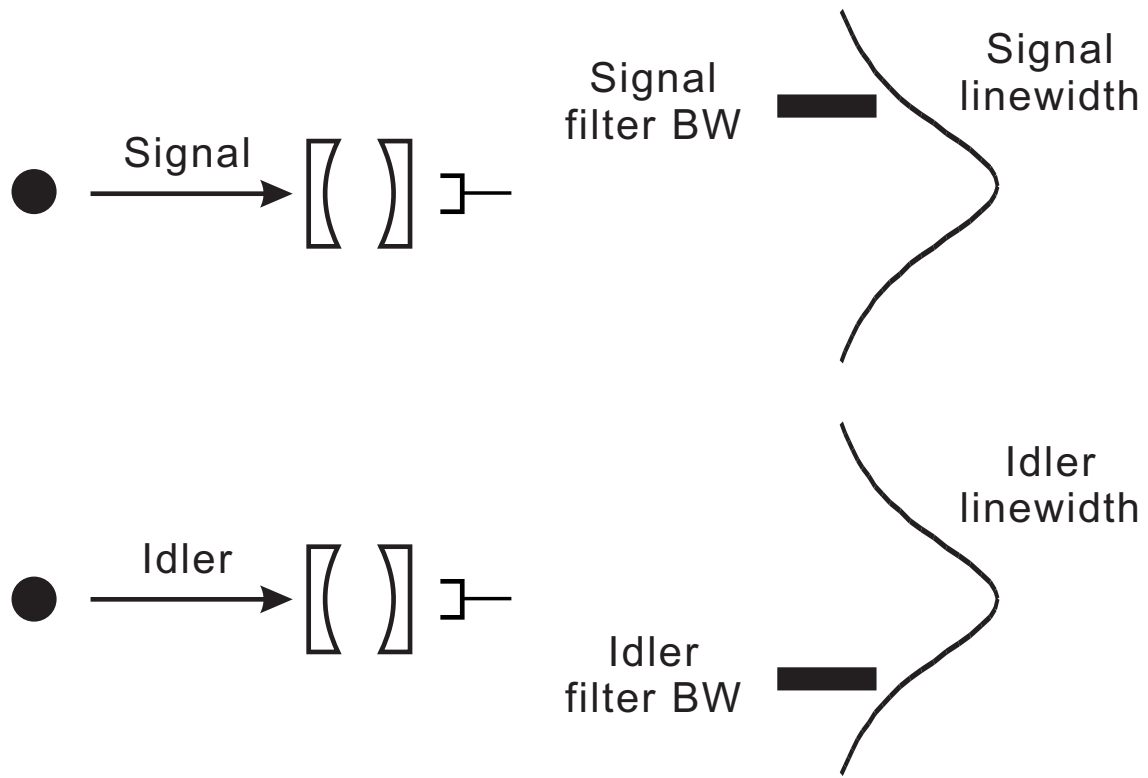,width=15cm}
\end{center}
\caption{Schematic for filter-penetration optical 
magic bullets.  BW: bandwidth.}
\end{figure}

\end{document}